\documentclass{jpsj-suppl}
\usepackage{wrapfig}
\title{Update of HKN Nuclear PDFs}

\author{Masanori Hirai}

\inst{Nippon Institute of Technology, Saitama 345-8501, Japan.}

\email{mhirai@nit.ac.jp}

\recdate{February 28, 2016}

\abst{
We discuss consistency of the nuclear effects between the electromagnetic and weak interactions.
In order to study a possibility of different nuclear effects in the neutrino DIS process,
double differential cross section data are compared with
these values obtained by the HKN07 nuclear parton distribution functions (nPDFs).
Discrepancies are found in the small and large-$x$ regions, 
and difference of kinematical value $y$ dependence exists between the $\nu$ and $\bar{\nu}$ data around $x=0.35$.
Moreover, we study statistical significance of the neutrino DIS data in each $x$ bin
and discuss about the possibility of the different nuclear modifications.
}

\kword{neutrino, nucleus, parton, quark, gluon, QCD, DIS}

\begin{document}
\maketitle

\section{Introduction}
Experiments using neutrino beams are useful to understand weak interaction phenomena, 
and they gives effective information 
about flavor structure of parton distribution functions (PDFs)
by comparison with the experiments using charged lepton beams.
An energy flux of generated neutrino is not under control,
and it spreads into the wide energy range. 
Measured value,
for instance a scattered final lepton tagging,
could be integrated over of the incident neutrino energy $E_\nu$ as
\vspace{-2mm}
\begin{equation}
  \frac{d^2\sigma_{exp}}{d\Omega dE_l} \propto \int \frac{d^2 \sigma}{d\Omega dE_l}(E_{\nu})\Phi(E_{\nu}) dE_{\nu}.
\end{equation}
The energy flux $\Phi(E_{\nu})$ must be estimated by using an event generator.
However, details of scattering processes are different in the energy region,
which are quasi-elastic, resonance, and deeply inelastic scattering (DIS).
Therefore, the event generator should describ well these processes.
Due to weak interaction in the charged current (CC) process, 
the experiment for 
obtaining statistically enough number of events needs high density and thick targets,
and one generally uses nucleus targets.
Therefore, nuclear modifications should be considered,
so that it becomes important to establish the event generator 
for experiments covering the wide neutrino energy range,
especially MeV $\sim$ a few GeV.

\begin{wrapfigure}[17]{r}{80mm}
  \begin{center}
    \includegraphics[width=7.5cm]{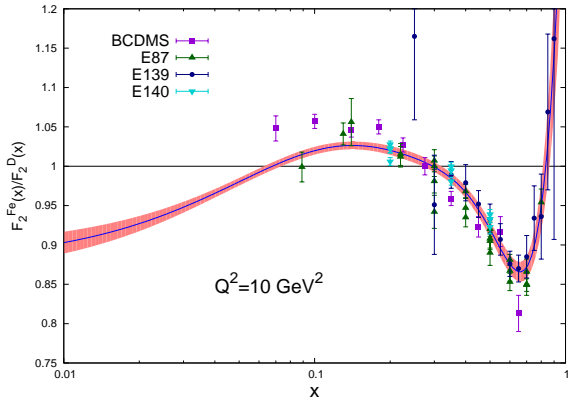}
  \end{center}

  \caption{The nuclear effects of the $F_2^A(x)$ for the Fe target.
           These experimental data have different $Q^2$ values.
           As a reference, curve and its uncertainty are obtained 
           by the HKN07 in NLO at $Q^2$=10 GeV$^2$ \cite{HKN07}.
           }
  \label{fig:R_Fe}
\end{wrapfigure}

As a high energy phenomenon, 
the DIS process is described well by the perturbative QCD (pQCD).
By charged lepton-nuclear DIS experiments, 
nuclear effects are measured as ratios
of the structure functions; $R^A(x)=F_2^A(x)/F_2^D(x)$.
The ratio of the Fe target is shown in Fig. \ref{fig:R_Fe}.
In the $x \le 0.08$ region, the suppression ($R^A<1$) is called the shadowing effect, 
the enhancement ($R^A>1$) in the range ($0.08<x \le 0.25$) is the anti-shadowing effect,
the suppression in the range ($0.25<x<0.8$) is the EMC effect,
and steeply rising behavior in the $x \ge 0.8$ region is the Fermi-motion effect.
Although quantitative behaviors of each effect are discussed by 
vector meson dominance, convolution model and so on,
it is desirable that these effects are estimated uniformly and qualitatively.
Therefore, PDFs including these effects are redefined as nuclear PDFs (nPDFs),
which are determined with the several experimental data by a global analysis
\cite{HKN07,EKS09,DSSZ12,nCETEQ15}.

So far the neutrino DIS experiments are performed by using the Fe and Pb targets \cite{NuTeV,CHORUS,CDHSW}.
A problem occurred in global analysis of nPDFs by using the data.
It is suggested that 
the nuclear effects of the CC process are different 
from that of the electromagnetic (EM) interaction process \cite{SYKMOO}.
Although properties of a bound nucleon could change in a nucleus, 
it is considered that 
the nuclear effects come from motion and structure of bound nucleons in a nucleus. 
For instance, 
they are described by a spectral functions of the nucleon in the convolution model.
Since these are intuitively based on the strong interaction, 
the effects would be independent on the probe of the EM and weak interactions.
This assumption is validated in the papers \cite{DSSZ12,PS}.

In the analysis \cite{DSSZ12}, 
the charged lepton and neutrino DIS, Drell-Yan (DY), and collider experimental data are used together. 
Moreover, structure functions (SFs) are used as fitting data of the CC DIS process.
In a global analysis, it is desirable that the fitting data are close to measurable value as much as possible.
Although about four thousand data exist as the differential cross section of the neutrino DIS experiments,
the number of data transformed into the SFs becomes an order of magnitude less. 
By applying additional approximations in the transformation, 
there is possibility to lose significant information 
and statistical superiority.
In the $\chi^2$ analysis by combining data of several scattering processes, 
an estimated object must have universality.
It is therefore assumed that the nuclear effects are the same between the EM and CC reaction, 
whereas difference of the effects cannot be discussed in such global analysis.
In that sense, the analysis using the SFs data could not be sufficient verification.

Moreover, a larger number of the data has high priority for reproduction of them.
As this fact was suggested in the paper \cite{KSOYKMOS},
the number of the differential cross section data of the neutrino DIS is larger 
than that of the other scattering processes.
In order to adjust the priority, 
a weight is adopted for the total $\chi^2$ as a following;
$\chi^2_{\rm total}=\chi^2_{EM}+w\chi^2_{CC}$.
By the value of the weight $w$,
it is possible to obtain intermediate result describing the both data
even if the effects are different.
For the reason,  
modification of the nuclear effects must be evaluated by only using the neutrino DIS data.

Compatibility of the neutrino DIS data is discussed 
by comparing of the differential cross section \cite{PS}.
It is apparent that the nuclear effects must be required 
to describe qualitative behavior of the neutrino-nucleus DIS data
as shown in the $Q^2$ averaged figures.
However, the modification for the effects should be discussed 
by preforming the global analysis with the CC reaction data only.
It is premature to conclude that the consistency of the nuclear effects is kept
in the charged lepton and neutrino DIS precesses.

\section{Consistency check of neutrino DIS data}
\subsection{Neutrino DIS}
Since neutrino DIS experiments used rather heavy nucleus targets, 
it is difficult to show the $x$ dependence of the nuclear effect
as the ratio with a light nucleus.
As experimental data, we adopt the double differential cross section 
for the reason described in the above section.
The differential cross section of the neutrino DIS precess can be expressed by 
\begin{equation}
  \frac{1}{E_{\nu}}\frac{d^2\sigma^{CC}}{dxdy} 
    = \frac{G_F^2M_N}{\pi\left(1+\frac{Q^2}{M_W^2}\right)^2}
      \left[xy^2F_1^{CC}+\left( 1-y-\frac{(M_Nxy)^2}{Q^2}\right)F_2^{CC}
            \pm xy\left(1-\frac{y}{2}\right)F_3^{CC})
      \right],
\end{equation}
where $G_F$ indicats the Fermi coupling constant,
$M_W$ is the $W$ boson mass. 
$M_N$ is nucleon mass and is changed to averaged nucleon mass in a nuclear target case.
$x$ is Bjorken variable, and $y$, $E_\nu$ are kinematical values. 
On the right side of the equation, 
incident neutrino energy dependence is included into the definition of $Q^2(=2M_NxyE_\nu$). 
$\pm$ indicates $+$ for neutrino, $-$ for anti-neutrino beam case, respectively.
$F_1^{CC}$, $F_2^{CC}$, and $F_3^{CC}$ are structure functions for the CC interaction process,
and they have $x$ and $Q^2$ dependence.
These are obtained by following flavor combinations of PDFs.
\begin{equation}
\begin{aligned}[l]
  F_2^\nu         &= 2x(d+s+\bar{u}+\bar{c}), \ \  F_3^\nu         = x(d+s-\bar{u}-\bar{c}), \\
  F_2^{\bar{\nu}} &= 2x(u+c+\bar{d}+\bar{s}), \ \  F_3^{\bar{\nu}} = x(u+c-\bar{d}-\bar{s}).
  \label{eq:SF}
\end{aligned}
\end{equation}
The combination is different between the $\nu$ and $\bar{\nu}$ 
caused by charged current via the $W^{\pm}$ boson, 
therefore the neutrino DIS process is sensitive to flavor dependence for the PDFs.
As a simple method for a nuclear target case,
which includes the nuclear effects,
nPDFs are used instead of the PDFs.
The nPDFs are determined by using experimental data of several nuclear targets,
and so these would not depend on theoretical models.

In our analysis, nPDF is defined as a parton distribution in a nucleus
and expressed by using weight functional form:
\begin{equation}
  f_i^A(x) \equiv w_i(x,A,Z)\frac{\left[Zf_i^p(x)+(A-Z)f_i^n\right]}{A}, 
  \label{eq:HKN07}
\end{equation}
where $i$ means $q$, $\bar{q}$, and gluon.
$f_i^{p,n}$ indicates PDF in a free nucleon, 
and isospin symmetry is assumed: $f_{u}^p=f_{d}^n$, $f_{d}^p=f_{u}^n$.
A, Z indicate mass and charged numbers of a target nucleus, respectively. 
As a weight function $w_i(x,A,Z)$, 
a cubic function is adopted to reproduce the nuclear effects
in whole $x$ region \cite{HKN07}.
Baryon, charge-number, and momentum conservation are satisfied 
by fixing a few parameters.
By regarding a nucleus as a free particle  
and redefining a nPDF in it,
factorization theorem would be kept. 
As an analogy for a free nucleon,
$Q^2$ dependence of nPDFs can be obtained by the DGLAP equation.
The nPDFs as including nuclear effects are non-perturbation part 
which must be determined with experimental data.

We noted that
a following definition is used
by other analysis groups \cite{EKS09,DSSZ12,nCETEQ15}.
\begin{equation}
  f_i^A(x) \equiv \frac{Zf_i^{p/A}(x)+(A-Z)f_i^{n/A}(x)}{A}.
\end{equation}
$f_i^{p/A}(x)$ is regarded as a nPDF in a bound proton.
This interpretation is different in our analysis.
Although the same word ``nPDFs" is used, 
it is a confusing point that 
$f^A_i(x)$ and SFs are comparable in the both definitions;
however, definition of a ratio $R_i^A(x)$, which indicates the nuclear effects of each nPDF,
is different; $R_i^A(x)=f_i^A(x)/f_i^p(x)$ in the HKN07
and $R_i^A(x)=f_i^{p/A}(x)/f_i^p(x)$, for example \cite{nCETEQ15}.
It therefore must be taken care when comparing them in such figures
among these papers.

\begin{wrapfigure}[16]{r}{80mm}
  \begin{center}
    \includegraphics[width=7.5cm]{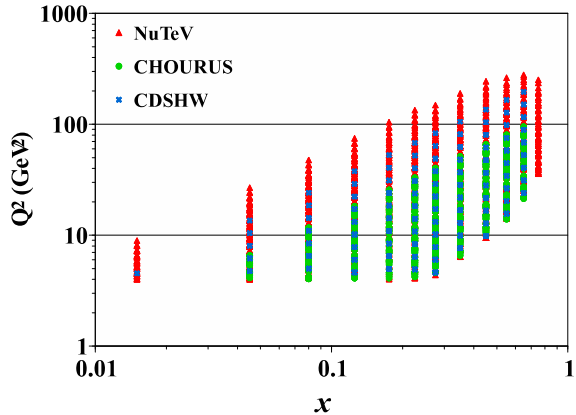}

    \caption{The kinematical range is shown by $x$ and $Q^2$ values 
             of each neutrino DIS experiment}
    \label{fig:XvsQ2}
  \end{center}
\end{wrapfigure}

The kinematical range of the neutrino DIS data is shown in Fig. \ref{fig:XvsQ2},
and number of experimental data are shown in Table \ref{tab:data}.
For eliminating higher order and twist effects,
we apply the following kinematical cut; 
$Q^2>4$ GeV$^2$, $W>3.5$ GeV.
The square of invariant mass is defined as a following; 
$W^2=M_N^2+2M_N E_\nu y(1-x)$.
The experimental measurements are shown by $x$, $y$, and $E_\nu$ bins.
The data are excluded 
in the lower-$x$ and $E_\nu$ bins by the kinematical cut of the $Q^2$ value,
and also in the large-$x$ and lower-$E_\nu$ bins by the $W$ cut.

\begin{table}[t]
\begin{center}
\caption{Number of the data for the neutrino DIS.
        Values of the $\chi^2$ per the number without and with normalization factors.
        }
\label{tab:data}
\begin{tabular}{cccccccc}
\hline
\raisebox{-1.8ex}[0pt][0pt]{Experiment} & \raisebox{-1.8ex}[0pt][0pt]{Reference} & \raisebox{-1.8ex}[0pt][0pt]{Target} 
 & \multicolumn{2}{c}{\# of data} & \multicolumn{2}{c}{$\chi_{\nu+\bar{\nu}}^2/($\# of data)} & \raisebox{-1.8ex}[0pt][0pt]{norm} \\
 &  &  & $\nu$ & $\bar{\nu}$ & without norm & with norm &  \\
\hline\hline
NuTeV      & \cite{NuTeV}  & Fe   & 1168 & 966 & 1.54 & 1.23 & 0.96 \\
CHORUS     & \cite{CHORUS} & Pb   &  412 & 412 & 1.58 & 0.96 & 0.94 \\
CDHSW      & \cite{CDHSW}  & Fe   &  465 & 464 & 1.18 & 0.91 & 0.96 \\
\hline
\end{tabular}
\end{center}
\end{table}

These data cover four nuclear effects regions:
$0.015 <x \le 0.08$ (shadowing effect), 
$0.08 <x  \le 0.25$ (anti-shadowing effect),
$0.25 <x < 0.8$ (EMC effect),
$x \ge 0.8$ (Fermi-motion effect).
Most data exist in the anti-shadowing and EMC regions,
and these data expect to improve the nuclear effects on the valence-quark distribution
which becomes the main component in the medium-$x$ region.
Although the data in the small-$x$ region could affect on 
the determination of the sea-quark distribution, 
which is the main contribution of the shadowing effect,
it is difficult to obtain enough accuracy 
from rather small number of the data
restricted by the kinematical cut.

\subsection{Comparison of the neutrino DIS data}
\begin{figure}[th!]
  \begin{center}
    \includegraphics[width=15cm]{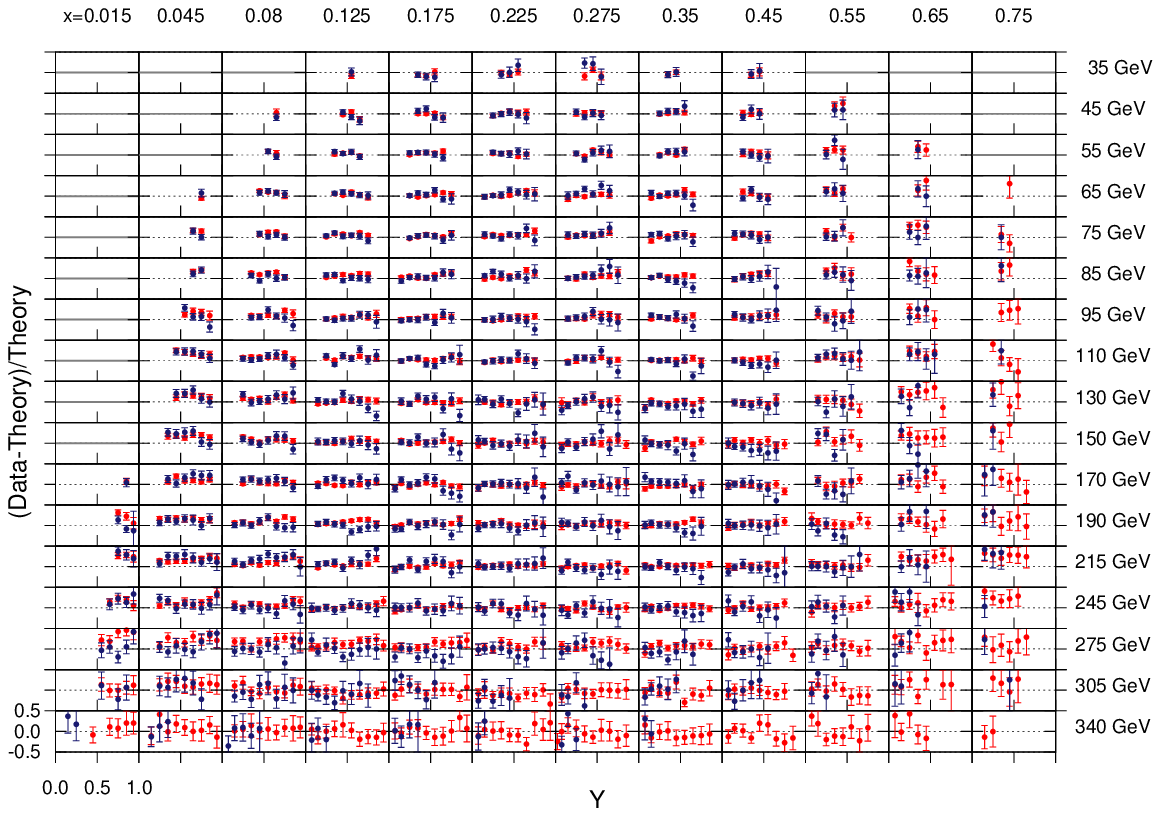}
  \end{center}

  \caption{Comparison of the NuTeV experimental data for different $x$ and $E_{\nu}$ bins.
           Red circle indicates neutrino, and blue circle is anti-neutrino. 
           The abscissa of each bin is kinematical value y.
           }
  \label{fig:NuTeV}
  \begin{center}
    \includegraphics[width=15cm]{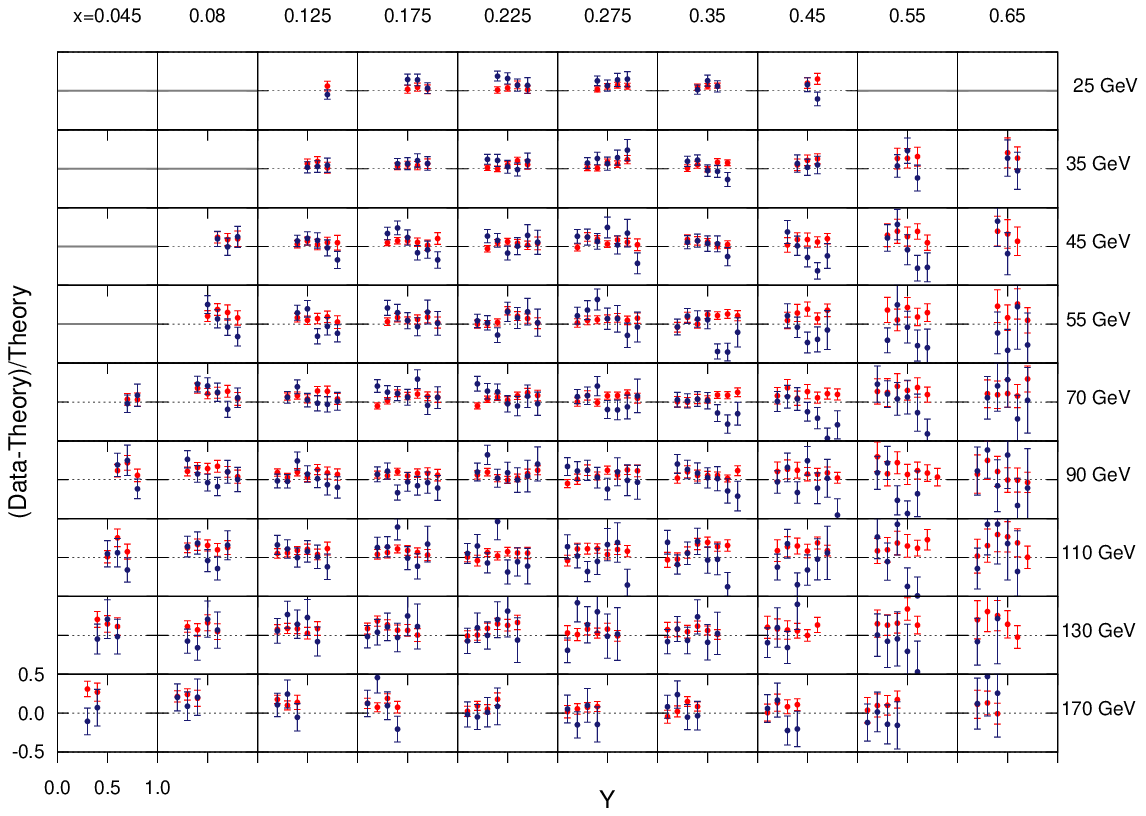}
  \end{center}

  \caption{Comparison of the CHORUS experimental data for different $x$ and $E_{\nu}$ bins.}
  \label{fig:CHORUS}
\end{figure}

\begin{figure}[t!]
  \begin{center}
    \includegraphics[width=15cm]{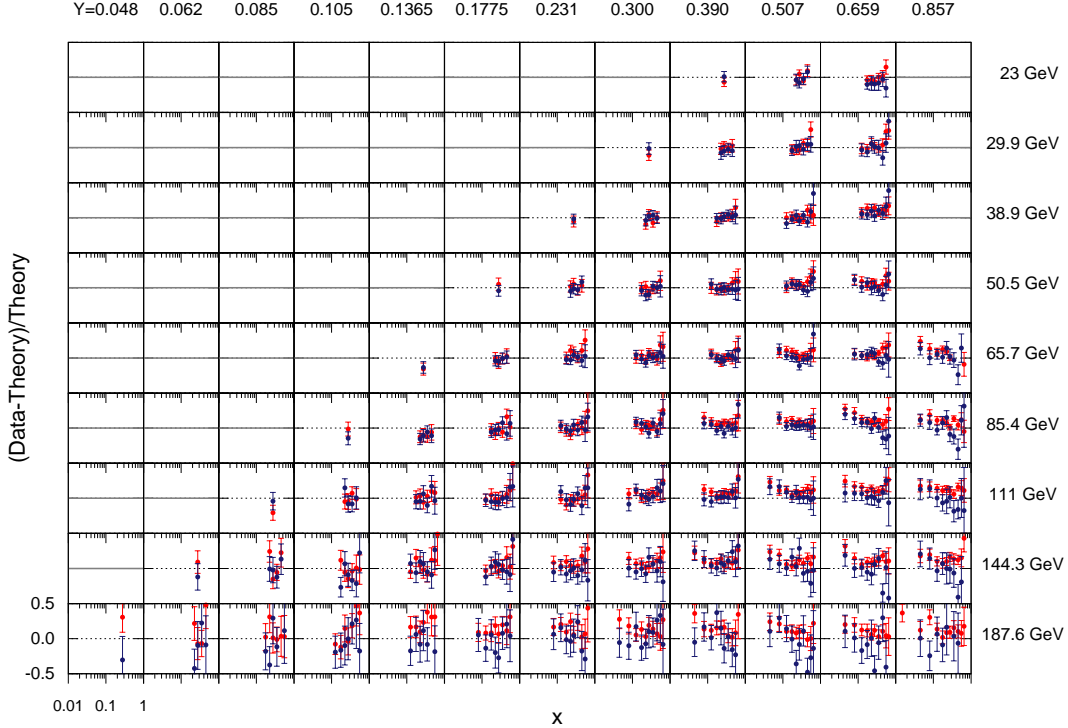}
  \end{center}
  \caption{Comparison of the CDHSW experimental data for different $y$ and $E_{\nu}$ bins.
           The abscissa of each bin shows in the range $0.01<x<1$.}
  \label{fig:CDHSW}
  \vspace{-5mm}
\end{figure}

In order to check consistency of the neutrino DIS data,
double differential cross sections are calculated by using the HKN07 nPDFs at NLO.
The nPDFs are obtained by using the EM DIS and Drell-Yan data of several nuclear targets \cite{HKN07}.
The comparison with the neutrino data are shown in Figs. \ref{fig:NuTeV}, \ref{fig:CHORUS}, and \ref{fig:CDHSW}.
For clarifying relative difference between the data and theoretical values 
including the nuclear effects of the EM reaction, 
this figure plots the rational difference: (Data-Theory)/Theory.

From the NuTeV experiment data in Fig. \ref{fig:NuTeV}, 
we can find discrepancies in the $x=0.015$ and $0.045$ columns.
It would indicate that more shallow shadowing effect is required for fitting to the data.
In fact, the shallower shadowing and suppression of the antishadowing effect 
are suggested in the paper \cite{SYKMOO}.
On the other hand, 
both results seem to be consistent in the region $0.125 \le x \le 0.55$ and $E_\nu \le 245$ GeV;
however, we can find difference of the $y$ dependence between $\nu$ and $\bar{\nu}$ data around $x=0.45$.
Although the nuclear effects, as qualitative behavior, are necessary to describe the data,
we cannot conclude that the effects are quantitatively the same as the EM interaction case.

The comparison of the CHORUS experiment data is shown in Fig. \ref{fig:CHORUS}.
Since the Pb nucleus is large neutron excess; A=208 and Z=82,
it is useful to discuss about flavor dependence of the nPDFs.
The flavor combinations of the nPDFs in the SFs are different between $\nu$ and $\bar{\nu}$ beam
as shown in Eq. (\ref{eq:SF}),
therefore the difference could be sensitive to 
the flavor dependence, especially up and down quark distributions.
We also find the significant difference of the $y$ dependence between $\nu$ and $\bar{\nu}$ 
in the $x=0.35$ and $0.45$ columns
where the valence-quark distributions are dominant.

\clearpage
The differences are rather small for the Fe target data
because of small neutron excess: A=56 and Z=26.
Since the same nuclear effects on the valence quark distributions are assumed in the nPDF analysis,
which means $w_{u_v}^A=w_{d_v}^A$ in Eq. (\ref{eq:HKN07}),
the $u_v^{Fe}(x)$ and $d_v^{Fe}(x)$ become almost the same distributions.
By contrast, the difference between $u_v^{Pb}(x)$ and $d_v^{Pb}(x)$ must be emphasized 
due to the large neutron excess.
Although the difference is produced by the neutron excess ($Z < A-Z$) in Eq. (\ref{eq:HKN07}),
the difference cannot be simply described.
It therefore suggests a possibility of flavor dependence of the modification factors: $w_{u_v}^A \ne w_{d_v}^A$.
However, the errors of the data are larger than those of the Fe target.
It is possible to reduce its $\chi^2$ value by introducing normalization factors
as an uncertainty for incident neutrino energy $E_\nu$.
It therefore is difficult to quantitatively determine the effects with enough accuracy.
More precise measurements are necessary for 
discussing about the flavor dependence of the nuclear effects, 
especially for valence-quark distributions.

The CDSHW experimental data are shown as $x$ dependence for each $y$ and $E_\nu$ bins
in Fig. \ref{fig:CDHSW}.
It seems that discrepancies have $x$ dependence in the medium and large-$x$ regions.
The discrepancies appear significantly in $y=$0.390, 0.507 and $0.659$ columns.
Such rising behavior around $x=0.1$ clearly shows that the $x$ dependence of the nuclear effects 
is different between the EM and weak interaction processes.
The discrepancy is also found in the smaller-$x$ region ($x \le 0.08$) for other experiments.
Although these data in the region depends on kinematical cut $Q^2$ and $W$,
such behavior indicates that 
these data cannot be completely fitted by only including normalization factors  
and are expected to affect the modification of the $x$ dependence of the nPDFs.
We could not exclude the possibility of probe dependence of the nuclear effects,
especially the shadowing effect.
In the large-$x$ region, the inconsistency can be found in the NuTeV data at $x=0.65$ and $0.75$ columns
in Fig. \ref{fig:NuTeV}.
The Fermi-motion effect of nPDFs cannot be determined well 
due to lack of data with enough accuracy.
There is a possibility of improving the behavior 
with precise measurements of the EM reactions.

Finally, the errors of the neutrino data are rather large in the higher-$E_\nu$ range.
In the $\chi^2$ analysis, 
Large value of the $\chi^2$ becomes dominant contribution and has high priority.
In addition, the value is proportional to the number of data $N$, 
and generally goodness of fit is evaluated by $\chi^2/N \sim 1$.
However, these data become numerical noise in total $\chi^2$ value,
sensitivity for determination of the nuclear effects is reduced.
Therefore, a number of effective data could turn out to be small 
in spite of a large amount of existing neutrino data,
This fact should be noted 
when data sets, being significantly different number, are treated in the global analysis.

\subsection{Significance of neutrino DIS data}
As mentioned in the introduction, significance of the neutrino data must be independently investigated
for taking into account of the possibility that the nuclear effects are different between the EM and CC reactions.
As the theoretical values, the cross sections are calculated by using HKN07 at NLO.
The obtained values of the $\chi^2/N$ are shown in Table \ref{tab:data}.
These values are not so large
because the values of the charged lepton nucleus DIS data for Fe target are also the same as the HKN07:
$\chi^2_{\ell \rm{Fe-DIS}}/N=1.52$.
Due to uncertainty of an energy flux estimation in the neutrino experiments and 
restriction to a degree of freedom by applying functional form in the nPDFs analysis,
we adopt normalization factors which are obtained by fixing the nPDFs.
The uncertainty of these obtained factors is about 2\%.
The results are shown in the same table.
Although the details of the behaviors for each $x$, $y$, and $E_\nu$ bins cannot be reconstructed,
it seems to be good fitting results as long as they are evaluated by these values, 
which are, however, not at minimized point by allowing change of the $x$ dependence on the nPDFs.

\begin{wrapfigure}[15]{r}{75mm}
  \vspace{-13mm}
  \begin{center}
    \includegraphics[width=75mm]{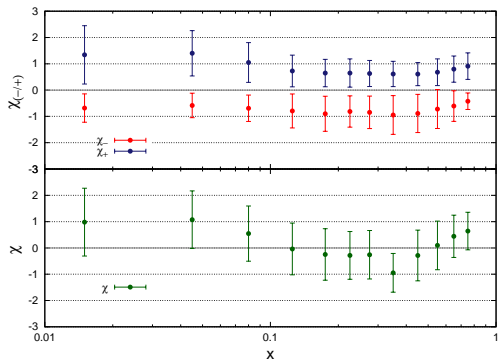}
  \end{center}

  \caption{Mean value of the $\chi$ distribution at each $x$ bin.  
           Error bar indicates its standard deviation of one.
          }
  \label{fig:chi_dist}
\end{wrapfigure}
In order to investigate the dependence,
we check contribution of the data in each $x$ bin.
For clarifying modification of the data,
$\chi$ distribution is studied.
Values of the $\chi$ are calculated by
  $\chi_i=\frac{D_i-T_i/norm_j}{\sigma_i}$,
where $i$ indicates the $x$ bin,
and $\sigma_i$ is experimental error is  corresponding to the data $D_i$.
$norm_j$ is a normalization factor of a data set of $j$.
Positive value of the $\chi$ means that
the data affect as upper modification on the theoretical value at $x$,
and negative is as downer one.
These $\chi_i$ in the different $y$ and $E_\nu$ bins distribute on a plane at $x$ indicated by the index of $i$.
Mean value of the $\chi$ distribution and standard deviation (SD) of one at each $x$ are shown 
in Fig. \ref{fig:chi_dist}.
$\chi_{+(-)}$ means taking only positive (negative) value of the $\chi_i$
in the upper figure.
The mean value and its SD are equivalent between them, 
which means that these data uniformly distribute around the theoretical value.
Since statistical difference cannot be found in the $0.175 \le x \le 0.275$ bins,
these data do not contribute to the modification of the current nPDFs.
By the same reason, the deepest point of the EMC effect does not almost change around $x=0.65$,
which is consistent with the fact that 
the EMC effect can be interpreted as a binding effect based on the strong interaction.
In the large-$x$ region, we cannot make clear discussion 
due to an inaccurate determination of the nPDFs themselves.
Disparity between the $\chi_+$ and $\chi_-$ can be found in the small-$x$ region ($x \le 0.08$) and at $x=0.35$.
Direction of the modification becomes apparent in the lower figure 
which indicates superposition of the $\chi_\pm$ at each $x$ bin.
Inconsistencies remain even if we adopt normalization factor.
Although statistical significance cannot be found in the data at the small $x$,
this fact indicates that shallower shadowing effect is required for the best-fitting to these data.
It is considered that different shadowing effect can be allowed due to the difference of the coupling
between vector and axial-vector mesons.
Moreover, the statistical difference at $x=0.35$ is caused by 
discrepancy of the behavior between the $\nu$ and $\bar{\nu}$ data
as shown in Figs. \ref{fig:NuTeV}, and \ref{fig:CHORUS}, 
whereas total contribution in the $\chi^2$ analysis seems to be a cause of deeper EMC effect than 
that of the EM interaction case.
Therefore, we cannot still conclude that it is evidence for the probe dependence of the nuclear effects.

\section{Summary}
Consistency of the neutrino data is discussed by comparing the cross section data with 
the theoretical values calculated by the HKN07 nPDFs
which obtained with the EM reaction data of several nuclear targets.
Inconsistencies are found in the small and large-$x$ region, 
and the difference of the $y$ dependence between the $\nu$ and $\bar{\nu}$ data exists around $x=0.35$,
especially it becomes significant for the Pb target data.
Therefore, flavor dependence of the nuclear modification factor should be considered.
In addition, detail information about the antiquark is also important
because the contribution of these distributions is emphasized as shown in Eq. (\ref{eq:SF}).
The difference between $\bar{u}(x)$ and $\bar{d}(x)$ of the free nucleon PDFs even is not determined well in the region,
therefore it affects on determination of the nPDFs via Eq. (\ref{eq:HKN07}).

In order to discuss about the modification for the $x$ dependence of the nPDFs,
the $\chi$ distributions are studied at each $x$ bin.
As shown in lower panel in Fig. \ref{fig:chi_dist}, 
the modifications for nPDFs are required in the small-$x$ region and around $x=0.35$.
It means shallower shadowing effect and gradual slope for the EMC effect,
and then the different nuclear effects could be described by 
smoothly connecting these contributions from the depth of the EMC effect.
We are performing the global analysis with only the neutrino DIS data,
and reproduce similar modification for the $F_2^{\ell \rm Fe}(x)$ suggested in the papers \cite{SYKMOO,KSOYKMOS}.
However, there is still room for consideration about obtained effects on each $f_i^A(x)$.
Since statistical significance is not enough to discuss about more detail on the issue,
we cannot conclude that there is the probe dependence of the nuclear effects.
Further studies are needed by using more precise neutrino DIS data 
and comparing with another scattering process, for example
$W$ production process in  pA collision by the RHIC and LHC experiments.

\section{Acknowledgments}
\vspace{-2mm}
The author thanks local organizers of the NuInt15 workshop
for their financial support. 


\end{document}